\documentclass{ifacconf}

\usepackage{natbib,graphicx}
\usepackage{amsmath}
\usepackage[caption=false,font=footnotesize]{subfig}
\usepackage{psfrag}
\usepackage{amssymb}
\usepackage{latexsym}
\usepackage{subfig}
\usepackage{algorithm}
\usepackage{algorithmic}
\usepackage{color}

\newtheorem{remark}{\bf Remark}[section]
\newtheorem{proposition}{\bf Proposition}[section]

\newtheorem{definition}{\bf Definition}[section]
\newtheorem{example}{\bf Example}[section]
\newtheorem{theorem}{\bf Theorem}[section]
\newtheorem{problem}{\bf Problem}[section]

\newcommand{\B}{\mathcal{B}}

\newcommand{\Prod}{\mathcal{A}}
\newcommand{\T}{\mathcal{T}}

\newcommand{\bx}{\textbf{x}}
\newcommand{\bq}{\textbf{q}}
\newcommand{\bp}{\textbf{p}}

\newcommand{\mP}{\mc{P}}
\newcommand{\bo}{\textbf{o}}

\newcommand{\Land}{\wedge}
\newcommand{\Lor}{\vee}
\newcommand{\Next}{\mathsf{X}\, }
\newcommand{\Always}{\mathsf{G} \,}
\newcommand{\Event}{\mathsf{F} \,}
\newcommand{\Until}{\, \mathsf{U} \,}

\newcommand{\st}{\,|\,}

\newcommand{\longto}{\longrightarrow}

\newcommand{\ssim}{\slash_{\sim}}

\newcommand{\inte}{\mathrm{int}}
\newcommand{\interior}{\mathrm{int}}

\newcommand{\cS}{\mathcal{S}}

\newcommand{\cP}{\mathcal{P}}

\newcommand{\co}{\mathrm{Co}}
\newcommand{\mx}{\mc X}
\newcommand{\mr}{\mc R}
\newcommand{\mt}{\mc D}
\newcommand{\rn}{\mb R^{n}}

\newcommand{\pre}{\mathrm{Pre}}
\newcommand{\eq}{\mathrm{eq}}
\newcommand{\tcp}{\tilde\cP}

\newcommand{\mc}[1]{\mathcal{#1}}
\newcommand{\mb}[1]{{\mathbb{#1}}}
\newcommand{\be}{\begin{equation}}
\newcommand{\ee}{\end{equation}}
\newcommand{\ben}{\begin{equation*}}
\newcommand{\een}{\end{equation*}}
\newcommand{\bea}{\begin{eqnarray}}
\newcommand{\eea}{\end{eqnarray}}
\newcommand{\bean}{\begin{eqnarray*}}
\newcommand{\eean}{\end{eqnarray*}}
\newcommand{\ba}{\begin{array}}
\newcommand{\ea}{\end{array}}
\newcommand{\leftm}{\left[\begin{array}}
\newcommand{\rightm}{\end{array}\right]}

\newcommand{\ie}{{\it i.e., }}
\newcommand{\eg}{{\it e.g., }}
\def\qed{\hfill\rule[-1pt]{5pt}{5pt}\par\medskip}

\definecolor{forestgreen}{rgb}{0,0.6,0.6}
\definecolor{brown}{rgb}{0.6,0.2,0.2}

\def\qed{\hfill\rule[-1pt]{5pt}{5pt}\par\medskip}

\newcommand\oprocendsymbol{\hbox{$\bullet$}}
\newcommand\oprocend{\relax\ifmmode\else\unskip\hfill\fi\oprocendsymbol}

\begin{document}
\begin{frontmatter}

\title{Formal Abstraction of Linear Systems via Polyhedral Lyapunov Functions\thanksref{footnoteinfo}} 

\thanks[footnoteinfo]{This work is partially supported by the ONR-MURI Award N00014-09-1051 at Boston University and the STW Veni Grant 10230 at Eindhoven University of Technology.}

\author[First]{Xuchu Ding}
\author[Second]{Mircea Lazar}
\author[Third]{Calin Belta}


\address[First]{Dept. of Systems at United Technologies Research Center, East Hartford, CT 06108, USA. Email: dingx@utrc.utc.com}
\address[Second]{Dept. of Electrical Engineering at Eindhoven University of Technology, Eindhoven, The Netherlands. Email: m.lazar@tue.nl}
\address[Third]{Dept. of Mechanical Engineering at Boston University, Brookline, MA 02446, USA.  Email: cbelta@bu.edu}

\begin{abstract}
In this paper we present an abstraction algorithm that produces a finite bisimulation quotient for an autonomous discrete-time linear system. We assume that the bisimulation quotient is required to preserve the observations over an arbitrary, finite number of polytopic subsets of the system state space.  We generate the bisimulation quotient with the aid of a sequence of contractive polytopic sublevel sets obtained via a polyhedral Lyapunov function.  The proposed algorithm guarantees that at iteration $i$, the bisimulation of the system within the $i$-th sublevel set of the Lyapunov function is completed.   We then show how to use the obtained bisimulation quotient to verify the system with respect to arbitrary Linear Temporal Logic formulas over the observed regions.
\end{abstract}
\end{frontmatter}

\section{Introduction}
\label{sec:intro}
In recent years, there has been a trend to bridge the gap between control theory and formal methods.   Control theory allows verifications of ``simple'' specifications (such as stability or reachability) for ``complex'' dynamical systems with a possibly infinite state space, while formal verification methods enable validation of a ``simple'' finite system in a ``complex'' (rich and expressive) specification language. Recent studies in the area of abstraction allow one to model the behaviors of complex dynamical systems as finite systems, so that formulas in a rich specification language such as Linear Temporal Logic (LTL) can be used to analyze, verify and control the behavior of the system, with applications in areas such as robotics \citep{belta2007symbolic}, multi-agent control systems \citep{Loizou04} and bioinformatics \citep{batt2005validation}.

In this paper, we focus on autonomous (without inputs) linear systems, and we aim to generate a finite bisimulation abstraction of the system within some relevant subset of the state space.  Since the bisimulation quotient preserves the language of the original infinite state system, it can be readily used for system verification. 

Our approach relies upon the existence of a \emph{polyhedral} Lyapunov function, which is non-conservative for stable linear systems, and we take advantage of the recent method by \cite{lazar2010infinity} to construct such Lyapunov functions.  The polyhedral Lyapunov function is used to generate a sequence of sublevel sets, which are contractive polytopes. We propose to partition the state space with respect to these polytopic sublevel sets, as they allow us to incrementally generate the bisimulation quotient of the entire relevant state space.  As the abstraction algorithm iterates, we guarantee that the bisimulation quotient is generated for an increasing larger sublevel set, with no ``holes'' in the covered state space.  The polytopic sublevel sets also ensure that the algorithm proposed in this paper only requires polytopic operations, and can be tractably implemented for systems in realistic and practical applications.

This work is related to relevant works on the construction of finite quotient for infinite systems, such as controlled linear systems \citep{tabuada2006linear, Pappas03} and hybrid systems \citep{Alur00}.  The bisimulation problem in general does not terminate \citep{Milner89}.  We side-step this issue by only considering the system behavior within a relevant state space,  i.e., in between two positive invariant compact sets that contain the origin in their interior.  Such positive invariant sets with arbitrary sizes can be immediately obtained from the polyhedral Lyapunov function as polytopic sublevel sets (\ie polytopes) centered at the origin.   Therefore, the bisimulation algorithm can capture any relevant subset of the state space.  This also directly gives a trade-off between the size of the bisimulation quotient and the size of the relevant state space being analyzed.

Another conceptually related work is \cite{sloth2010abstraction}, where two orthogonal (quadratic) Lyapunov functions were used for the abstraction of continuous-time Morse-Smale systems (including hyperbolic linear systems) to timed automata. Besides targeting general discrete-time linear systems, the main difference between \cite{sloth2010abstraction} and the approach proposed in this paper comes from the usage of \emph{polyhedral Lyapunov functions}. This turns out to be beneficial, as it removes the need for two orthogonal Lyapunov functions and it results in a tractable implementation.

The rest of the paper is organized as follows.  We introduce preliminaries in Sec.~\ref{sec:prelim} and formulate the problem in Sec.~\ref{sec:problem}.  We present the algorithm to generate the bisimulation quotient in Sec.~\ref{sec:genBis}, and we show in Sec.~\ref{sec:verification} how the resulting bisimulation quotient can be used to verify the system behavior against formulas in LTL.  Conclusions are summarized  in Sec.~\ref{sec:concl}.

\section{Preliminaries}
\label{sec:prelim}
For a set $\cS$, $\interior(\cS)$, $\partial(\cS)$, $\co(\cS)$, $|\cS|$, and $2^\cS$ stand for its interior, boundary, convex hull, cardinality, and power set, respectively. For $\lambda\in\Rset$ and $\cS\subseteq\Rset^n$, let $\lambda \cS:=\{\lambda x\st x\in \cS\}$. We use $\Rset,\:\Rset_+,\:\Zset,$ and $\Zset_+$ to denote the sets of real numbers, non-negative reals, integer numbers, and non-negative integers. For $m,n\in \Zset_+$, we use $\Rset^n$ and $\Rset^{m\times n}$ to denote the set of column vectors and matrices with $n$ and $m\times n$ real entries. 
For a vector $x\in\Rset^n$, $[x]_i$ denotes the $i$-th element of $x$ and $\|x\|_\infty=\max_{i=1,\ldots,n}\left|[x]_i\right|$ denotes the infinity norm of $x$. For a matrix $Z\in\Rset^{l\times n}$, 
let $\|Z\|_{\infty}:=\sup_{x\in \mathbb R^{n}\setminus 0}\frac{\|Zx\|_{\infty}}{\|x\|_{\infty}}$ denote its infinity norm. 


A $n$-dimensional \emph{polytope} $\mc P$ (see, \eg \cite{ziegler1995lectures}) in $\rn$ can be described as the convex hull of $n+1$ affinely independent points in $\rn$.  Alternatively, $\mc P$ can be described as the intersection of $k$, where $k\geq n+1$, closed half spaces, \ie there exists $k\geq n+1$ and $H_{\mc P}\in \mb R^{k\times n}$, $h_{\mc P}\in \mb R^{k}$, such that
\be
\label{eq:definitionofpolytope}
\mc P=\{x\in \rn \st H_{\mc P}x\leq h_{\mc P}\}.
\ee
We assume polytopes in $\rn$ are $n$-dimensional unless noted otherwise.
The set of boundaries of a polytope $\cP$ are called \emph{facets}, denoted by $f(\cP)$, which are themselves $(n-1)$-dimensional polytopes.  A \emph{semi-linear} set (sometimes called a \emph{polyhedron} in literature) in $\rn$ is defined as finite unions, intersections and complements of sets
$\{x\in \rn \st a^{\mathtt{T}}x\sim b, \sim \in \{=, <\}\}$, for some $a\in \rn$ and $b\in \mb R$.  Note that a convex and bounded semi-linear set is equivallent to a polytope with some of its facet removed.


\subsection{Transition systems and bisimulations}
\label{sec:tsandbisim}
\begin{definition}
\label{def:tran_sys}
A transition system (TS) is a tuple $\T=(Q,\to,\Pi,h)$, where
\begin{itemize}
\item $Q$ is a (possibly infinite) set of states;
\item $\to \subseteq Q\times Q$ is the set of transitions;
\item $\Pi$ is a finite set of observations; and
\item $h : Q \longrightarrow 2^{\Pi}$ is the observation map.
\end{itemize}
We denote $x\to x'$ if $(x,x')\in\to$.   We assume $\mathcal T$ to be non-blocking, \ie for each $x\in Q$, there exists $x'\in Q$ such that $x\to x'$. A {\em trajectory} of a TS from an initial state $x_{0}$ is an infinite sequence $\bx=x_{0}x_{1}...$ where $x_{k}\to x_{k+1}$ for all $k\in \mb Z_{+}$.  A trajectory $\bx$ generates a word $\bo=o_{0}o_{1}...$, where $o_{k}=h(x_{k})$ for all $k\in \mb Z_{+}$.   
\end{definition}
The TS $\T$ is \emph{finite} if $|Q|<\infty$, otherwise $\T$ is \emph{infinite}.  Moreover, $\T$ is \emph{deterministic} if for all $x\in Q$, there exists at most one $x'\in Q$ such that $x\to x'$, otherwise, $\T$ is called \emph{non-deterministic}. Given a set $X\subseteq Q$, we define:
\be
\label{eq:pre}
\mathrm{Pre}_\T (X)=\{x\in Q \st \exists x'\in X, x\to x'\},
\ee
\ie $\mathrm{Pre}_\T(X)$ is the subset of $Q$ that reaches $X$ in one step.  At a state $x\in Q$, the set of all words generated by trajectories originating from $x$ is called the \emph{language} of $\T$ originating at $x$, which is denoted by $\mc L_{\T}(x)$.  We also denote by $\mc L_{\T}(X)$ the language of $\T$ originating from states in a subset $X\subseteq Q$.

States of a TS can be related by a relation $\sim\subseteq Q\times Q$.  For convenience of notation, we denote $x\sim x'$ if $(x,x')\in \sim$.   The subset $X\subseteq Q$ is called an \emph{equivalent class} if $x,x'\in X \Leftrightarrow x\sim x'$.  We denote by $Q/_{\sim}$ the set labeling all equivalent classes and define a map $\eq:Q/_{\sim}\longrightarrow 2^Q$ such that $\eq(X_\sim)$ is the set of states in the equivalence class $X_\sim\in Q\ssim$.
\begin{definition}
\label{def:opr}
We say that a relation $\sim$ is \emph{observation preserving} if for any $x,x'\in Q$, $x\sim x'$ implies that $h(x)=h(x')$.
\end{definition}

A finite \emph{partition} $P$ of a set $\mc S$ is a finite collection of sets $P:=\{P_{i}\}_{i\in I}$, such that $\cup_{i\in I}P_{i}=\mc S$ and $P_{i}\cap P_{j}=\emptyset$ if $i\neq j$.  A finite \emph{refinement} of $P$ is a finite partition $P'$ of $\mc S$ such that for each $P_{i}\in P'$, there exists $P_{j}\in P$ such that $P_{i}\subseteq P_{j}$. Note that $P$ is a trivial refinement of itself.

A partition naturally induces a relation, and an observation preserving relation induces a quotient TS.
Given a TS $\T=(Q,\to,\Pi,h)$, a partition $P$ of $Q$ induces a relation $\sim$, such that $x\sim x'$ if and only if there exists $P_{i}\in P$ and $x,x'\in P_{i}$.  If $\sim$ induced by $P$ is observational preserving, then $P$ is said to be an observation preserving partition.  One can immediately verify that a refinement of an observation preserving partition is also observation preserving.
\begin{definition}
Given a TS $\T=(Q,\to,\Pi,h)$ and an observation preserving relation $\sim$, a quotient transition system $\T\ssim=(Q\ssim,\to_{\sim},\Pi,h_{\sim})$ is a transition system, where
\begin{itemize}
 \item $Q/_{\sim}$ is the set labeling all equivalent classes;
 \item $\to_{\sim}$ is defined as follows:  given $X_\sim, Y_\sim\in Q\ssim$, $X_\sim\to_{\sim}Y_\sim$ if and only if there exists $x\in \eq(X_\sim)$ and $x'\in \eq(Y_\sim)$ such that $x\to x'$;
 \item The set of observations $\Pi$ is inherited from $\T$;
 \item $h_{\sim}(X_\sim):=h(x)$, where $x\in \eq(X_\sim)$ (note that this map is only well-defined if $\sim$ is observation preserving).
\end{itemize}
\end{definition}

\begin{definition}
Given a TS $\T=(Q,\to,\Pi,h)$, a relation $\sim$ is a bisimulation relation of $\T$ if (1) $\sim$ is observation preserving; and (2) for any $x_{1},x_{2}\in Q$, if $x_{1}\sim x_{2}$ and $x_{1}\to x_{1}'$, then there exists $x_{2}'\in Q$ such that $x_{2}\to x_{2}'$ and $x_{1}'\sim x_{2}'$.
\end{definition}
If $\sim$ is a bisimulation, then the quotient transition system $\T\ssim$ is called a \emph{bisimulation quotient} of $\T$.  In this case, $\T$ and $\T\ssim$ are said to be \emph{bisimilar}.  Bisimulation is a very strong equivalence relation between systems.  In particular, it implies language equivalence.  Specifically, we have
\be
\forall X_\sim\in Q\ssim,\quad \mc L_{\T}(\eq(X_\sim))=\mc L_{\T\ssim}(X_\sim).
\ee
This fact (see \citep{Milner89, browne1988characterizing,davoren2000logics}) ensures that a bisimulation preserves properties expressed in temporal logics such as LTL, CTL and $\mu$-calculus.   As such, it is used as an important tool to reduce the complexity of system analysis or verification, since the bisimulation quotient (which may be finite and thus much smaller) can be analyzed or verified instead of the original system.

\subsection{Polyhedral Lyapunov functions}
\label{sec:sec:lyapunovfun}
Consider an autonomous discrete-time system,
\be
\label{eq:dynDisSys}
x_{k+1}=\Phi(x_{k}),\quad k\in \mb Z_{+},
\ee
where $x_{k}\in \rn$ is the state at the discrete-time instant $k$ and $\Phi: \rn\rightarrow \rn$ is an arbitrary map with $\Phi(0)=0$.  Given a state $x\in\rn$, then $x':=\Phi(x)$ is a called the \emph{successor} state of $x$.



A function $\phi:\mb R_{+}\longto \mb R_{+}$ belongs to class $\mc K_{\infty}$ if if it is continuous, strictly increasing, $\phi(0)=0$ and $\lim_{s\to\infty}\phi(s)=\infty$.
\begin{definition}
\label{defPI}
We call a set $\cP\subseteq\Rset^n$ \emph{positively invariant} for system \eqref{eq:dynDisSys} if for all $x\in\cP$ it holds that $\Phi(x)\in\cP$.
Let $\lambda\in[0,1]$. We call $\cP\subseteq\Rset^n$ \emph{$\lambda$-contractive} (or shortly, contractive) if for all $x\in\cP$ it holds that $\Phi(x)\in\lambda\cP$. 
\end{definition}

The proof for the following theorem can be found in \citep{jiang2002converse,lazarThesis}.

\begin{theorem}
\label{def:lf}
Let $\mx$ be a positively invariant set for \eqref{eq:dynDisSys} with $0\in \inte(\mx)$.  Furthermore, let $\alpha_{1},\alpha_{2}\in \mc K_{\infty}$, $\rho\in(0,1)$ and $V:\rn\longto \mb R_{+}$ such that:
\be
\label{eq:LFcond1}
\alpha_{1}(\|x\|)\leq V(x)\leq\alpha_{2}(\|x\|), \forall x\in \mx,
\ee
\be
\label{eq:LFcond2}
V(\Phi(x))\leq\rho V(x), \forall x\in \mx.
\ee
Then system \eqref{eq:dynDisSys} is asymptotically stable in $\mx$.
\end{theorem}
\begin{definition} A function $V:\rn\rightarrow \mb R_{+}$ is called a \emph{Lyapunov function} (LF) in $\mx$ if it satisfies \eqref{eq:LFcond1} and \eqref{eq:LFcond2}.  If $\mx=\rn$, then $V$ is called a \emph{global Lyapunov function}.
\end{definition}
The parameter $\rho$ is called the \emph{contraction rate} of $V$. For any $\Gamma>0$, $\mc P_{\Gamma}:=\{ x\in \rn\st V(x)\leq \Gamma\}$ is called a \emph{sublevel set} of $V$.

For the remainder of this paper we consider LFs defined using the infinity norm, i.e.,
\be
\label{eq:polyLF}
V(x)=\|Lx\|_{\infty},\quad L\in \mb R^{l\times n}, l \geq n,
\ee
where $L$ has full-column rank. Notice that infinity norm Lyapunov functions belong to a particular class of $0$-symmetric polyhedral Lyapunov functions. We opted for this type of function to simplify the exposition but in fact, the proposed abstraction method applies to general polyhedral Lyapunov functions defined by Minkowski (gauge) functions of polytopes in $\Rset^n$ with the origin in their interior.
\begin{proposition}
\label{prop:rhocontractive}
Suppose that $L\in \mb R^{l\times n}$ has full-column rank and $V$ as defined in \eqref{eq:polyLF} is a global LF for system \eqref{eq:dynDisSys} with contraction rate $\rho\in(0,1)$. Then for all $\Gamma>0$ it holds that $\mc P_\Gamma$ is a polytope and $0\in \inte(\mc P_{\Gamma})$. Moreover, if $\Phi(x)=Ax$ for some $A\in \mb R^{n\times n}$, then \emph{for all $\Gamma>0$} it holds that $\mc P_\Gamma$ is a $\rho$-contractive polytope for \eqref{eq:dynDisSys}.
\end{proposition}
The proof of the above result is a straightforward application of results in \citep{blanchini:1994, lazar2010infinity}.


\section{Problem Formulation}
\label{sec:problem}
In this paper, we consider autonomous discrete-time linear and time-invariant (LTI) systems, \ie
\be
\label{eq:linearDyn}
x_{k+1}=Ax_{k},\quad k\in\mb Z_+,
\ee
where $A\in\mb R^{n\times n}$ is a strictly stable (i.e., Schur) matrix.  In this paper, we assume that a global polyhedral Lyapunov function (LF) of the form \eqref{eq:polyLF} with contraction rate $\rho\in(0,1)$ is known for system \eqref{eq:linearDyn} (see Sec. \ref{sec:sec:lyapunovfun}). The algorithm proposed in \citep{lazar2010infinity} is employed to construct such a function with a desired contraction rate. 


Let $\mx$ be a polytope $\mx:=\{x\mid\|Lx\|_\infty\leq \Gamma_\mx\}$ and $\mt$ be a polytope $\mt:=\{x\mid\|Lx\|_\infty\leq \Gamma_\mt\}$, where $L$ corresponds to the polytopic LF \eqref{eq:polyLF} of system \eqref{eq:linearDyn} and we assume that $0<\Gamma_\mt<\Gamma_\mx$.  Note that $\mt\subset \mx$ and $0\in \inte(\mt)\subset \inte(\mx)$.  We call $\mx$ the working set and $\mt$ the target set.  We are interested in analyzing and verifying the behavior of the system within $\mx$ with respect to polytopic regions in the state space, until the target set $\mt$ is reached (since $\mt$ is positively invariant, the system trajectory will remain within $\mt$ after it reaches $\mt$). Note that we can pick $\Gamma_\mt$ arbitrarily small and $\Gamma_\mx$ arbitrary large so as to capture any compact relevant subset of $\rn$.

\begin{remark}
Our results can be extended to arbitrary positively invariant sets $\mx$ and $\mt$, i.e., not obtained as the sublevel sets of \eqref{eq:polyLF}.  We chose to work with sublevel sets of the given polyhedral LF for the simplicity of presentation, and because such LFs allows us to easily construct a polytopic positive invariant set of any size.
\end{remark}

We assume that there exists a set $\mc R$ of polytopes indexed by a finite set $R$, \ie $\mc R:=\{\mr_{i}\}_{i\in R}$, where $\mr_{i}\subseteq \mx\setminus\mt$ for all $i\in R$, and $\mr_{i}\cap \mr_{j} =\emptyset$ for any $i\neq j$.

\begin{example}
\label{ex:simpleEx}
Consider a system as in \eqref{eq:linearDyn} with $A=\left(\begin{array}{cc}0.65 & 0.32 \\ -0.42 & -0.92\end{array}\right)$.  A polyhedral Lyapunov function was constructed with the method in \citep{lazar2010infinity}, where,
\[L=\left(\begin{array}{cccc}-0.0625 & 0.6815 & 0.9947 & 0.9947 \\ 1 & 1 & 0.6868 & -0.0678\end{array}\right)^\mathtt{T},\]
and $\rho=0.94$.  We chose $\Gamma_\mx=10$ and $\Gamma_\mt=5.063$.  We show the polytope $\mx$, $\mt$, and a set of polytopes $\mc R$ in Fig.~\ref{fig:partition}.
\begin{figure}[ht]
\begin{center}
    \includegraphics[scale=.38]{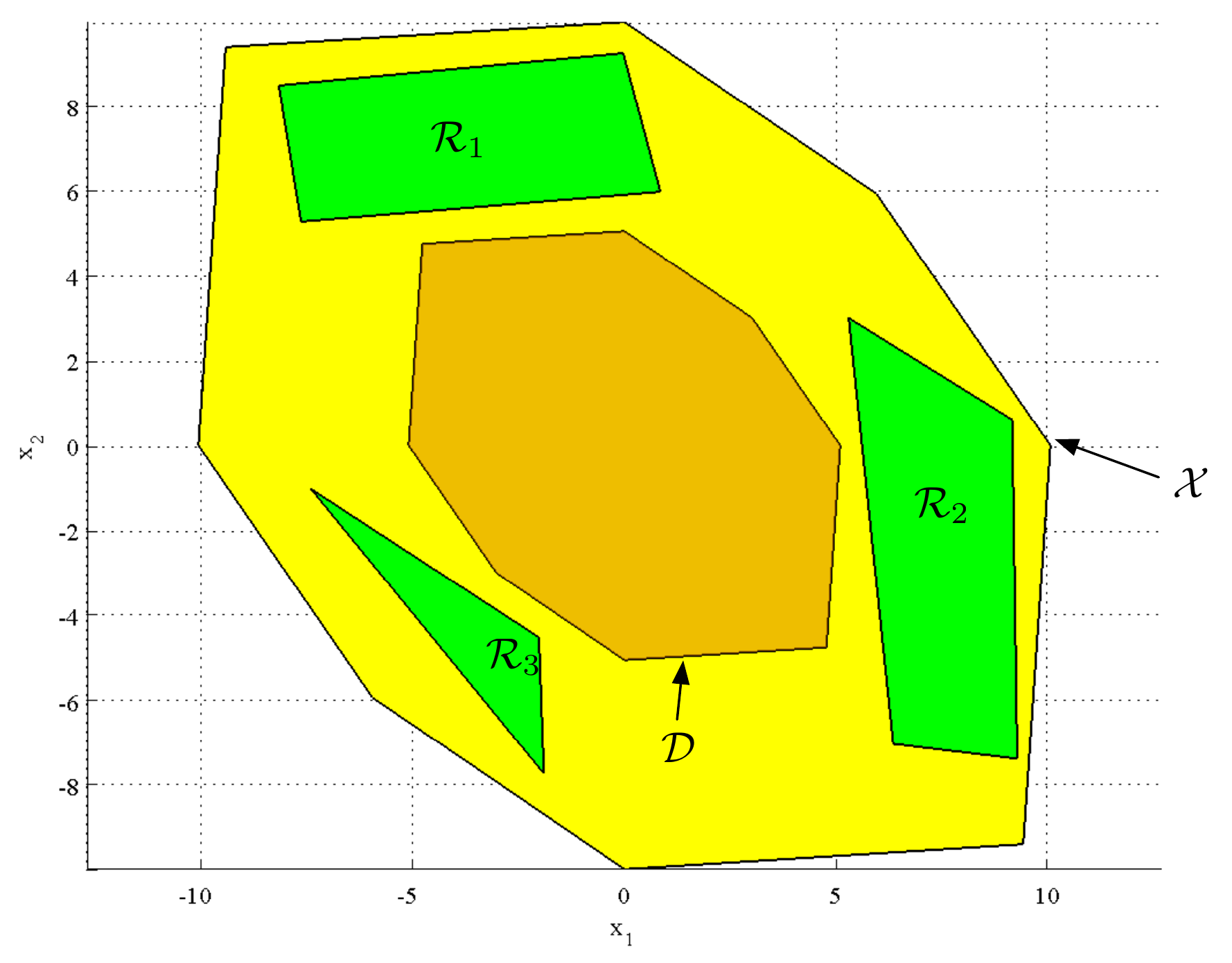}
   \caption{An example in $\mb R^2$ of the working set $\mx$ (in yellow), the target set $\mt$ (in brown), and a set of observational relevant polytopes $\mc R=\{\mr_1, \mr_2, \mr_{3}\}$ (in green).}
   \label{fig:partition}
\end{center}
\end{figure}
\end{example}

The set $\mc R$ represents regions of interest in the relevant state space, and the polytopes in $\mc R$ are considered as observations of \eqref{eq:linearDyn}.  Therefore, informally, a trajectory of \eqref{eq:linearDyn} $x_{0}x_{1}\ldots$ produces an infinite sequence of observations $o_{0}o_{1}\ldots$, such that $o_{i}$ is the index of the polytope in $\mc R$ visited by state $x_{k}$, or $o_{i}=\emptyset$ if $x_{k}$ is in none of the polytopes.  The definition of the semantics of the system can be formalized through an embedding of \eqref{eq:linearDyn} into a transition system, as follows.

\begin{definition}
\label{def:embeddedTS}
Let $\mx$, $\mt$, and $\mc R=\{\mr_i\}_{i\in R}$ be given.  The embedding transition system from \eqref{eq:linearDyn} is a transition system $\T_{e}=(Q_{e},\to_{e},\Pi_{e},h_{e})$ where
\begin{itemize}
\item $Q_{e}=\{x\in \rn \st x\in \mx\}$
\item \begin{enumerate}
\item If $x\in \mx\setminus\mt$, then  $x\to_{e}x'$ if and only if $x'=Ax$, \ie $x'$ is the state at the next time-step after applying the dynamics of \eqref{eq:linearDyn} at $x$;
\item If $x\in\mt$, $x\to_e x$ (since the target set $\mt$ is already reached, the behavior of the system after $\mt$ is reached is no longer relevant);
\end{enumerate}
\item $\Pi_{e}=R\cup \{\Pi_\mt\}$, \ie the set of observations is the set of labels of regions, plus the label $\Pi_\mt$ for $\mt$;
\item \begin{enumerate}
  \item $h_{e}(x):=i$ if and only if $x\in \mr_{i}$;
  \item $h_e(x):=\emptyset$ if and only if $x\in\mx\setminus(\mt\cup \bigcup_{i\in R}\mr_{i})$;
  \item $h_e(x):=\Pi_\mt$ if and only if $x\in \mt$.
\end{enumerate}
\end{itemize}
\end{definition}
Note that $\T_{e}$ is infinite and deterministic.  Moreover, $\T_{e}$ exactly captures the system dynamics under \eqref{eq:linearDyn} in the relevant state space $\mx\setminus\mt$, since a transition of the embedding TS $\T_{e}$ naturally corresponds to the evolution of the discrete-time system in one time-step (until the target set is reached).  Indeed, the trajectory of $\T_{e}$ from a state $x\in\mx\setminus\mt$ is exactly the same as the trajectory of the system from $x$ evolved under \eqref{eq:linearDyn} until $\mt$ is reached.  

%

The state space of $\T_e$ (which is the working set $\mx$) can be naturally partitioned as
\be
\label{eq:partition}
P_\mx:=\left\{\{\mr_i\}_{i\in R}, \mx\setminus(\mt\cup \bigcup_{i\in R}\mr_{i}), \mt\right\}.
\ee

%

It is straightforward to establish from the definition of $h_{e}$ in $\T_{e}$, that the relation induced from $P_\mx$ (see Sec. \ref{sec:tsandbisim}) is observation preserving.
We now formulate the main problem addressed in this paper.
\begin{problem}\label{prob:main}
Let a system \eqref{eq:linearDyn} with a polyhedral Lyapunov function of the form \eqref{eq:polyLF}, sets $\mx$, $\mt$ and $\{\mr_i\}_{i\in R}$ be given.  Find a finite observation preserving partition $P$ such that its induced relation $\sim$ is a bisimulation of the embedding transition system $\T_e$, and obtain the corresponding bisimulation quotient $\T_e\ssim$.
\end{problem}
\begin{remark}
 In fact, $P_{\mx}$ is the coarsest observation preserving partition for $\T_{e}$, and its induced relation is called an \emph{observation equivalence} relation in literature.  As a result, a finite partition is observation preserving if and only if it is a refinement of $P_{\mx}$.  Therefore, any solution of Prob. \ref{prob:main} is a refinement of $P_{\mx}$.
\end{remark}
\section{Generating the bisimulation quotient}
\label{sec:genBis}
Starting from a polyhedral Lyapunov function $V(x)=\|L x\|_{\infty}$ with a contraction rate $\rho=(0,1)$ as described in Sec. \ref{sec:sec:lyapunovfun} for system \eqref{eq:linearDyn}, we first generate a sequence of polytopic sublevel sets of the form $\mP_{\Gamma}:=\{x\in \rn \st \|Lx\|_{\infty}\leq \Gamma\}$ as follows.  Recall that $\mx=\mP_{\Gamma_\mx}$ and $\mt=\mP_{\Gamma_\mt}$ for some $0<\Gamma_\mt<\Gamma_\mx$.  We define a finite sequence ${\bar \Gamma}:=\Gamma_{0},\ldots,\Gamma_{N}$, where
\be
\label{eq:gammaseq}
\Gamma_{i+1}=\rho^{-1} \Gamma_{i}, \hspace{3mm} i=0,\ldots,N-2,
\ee
where $\Gamma_0:=\Gamma_\mt$, $\Gamma_{N}:= \Gamma_{\mx}$, and $N:=\arg\min_N\{\rho^{-N}\Gamma_0\mid \rho^{-N}\Gamma_0\geq \Gamma_\mx\}$.  The sequence $\bar\Gamma$ generates a sequence of sublevel sets $\bar{\mc P}_\Gamma:=P_{\Gamma_0},\ldots, P_{\Gamma_N}$. From the definition of the sublevel sets and $\bar\Gamma$, we have that
\be
\mP_{\Gamma_{0}}\subset\ldots \subset \mP_{\Gamma_{N}}.
\ee
Note that $\mc P_{\Gamma_0}$ is exactly $\mt$, $\mc P_{\Gamma_{N}}$ is exactly $\mx$, and $\mc P_{\Gamma_{N-1}}$ is the largest sublevel set defined via \eqref{eq:gammaseq} that is a subset of $\mx$.


Next, we define a \emph{slice} of the state space as follows:
\be
\label{eq:slices}
\mc S_{i}:=\mc P_{\Gamma_{i}}\setminus\mc P_{\Gamma_{i-1}}, \hspace{3mm} i=1,\ldots,N-1.
\ee
For convenience, we also denote $\mc S_{0}:=\mc P_{\Gamma_0}$ (although $\mc S_{0}$ is not a slice in between two sublevel sets).  We immediately see that the sets $\{\mc S_{i}\}_{i=0,\ldots,N}$ form a partition of $\mx$.  Note that the slices are bounded semi-linear sets (see Sec. \ref{sec:prelim}).

\begin{example}[Example \ref{ex:simpleEx} continued]
\label{ex:step2}
Consider the \\ system and sets as given in Example \ref{ex:simpleEx}.  The polytopic sublevel sets $\bar{\mc P}_\Gamma:=P_{\Gamma_0},\ldots, P_{\Gamma_N}$ are shown in Fig. \ref{fig:levelsets}.
\begin{figure}[h]
   \center
   \includegraphics[scale=.38]{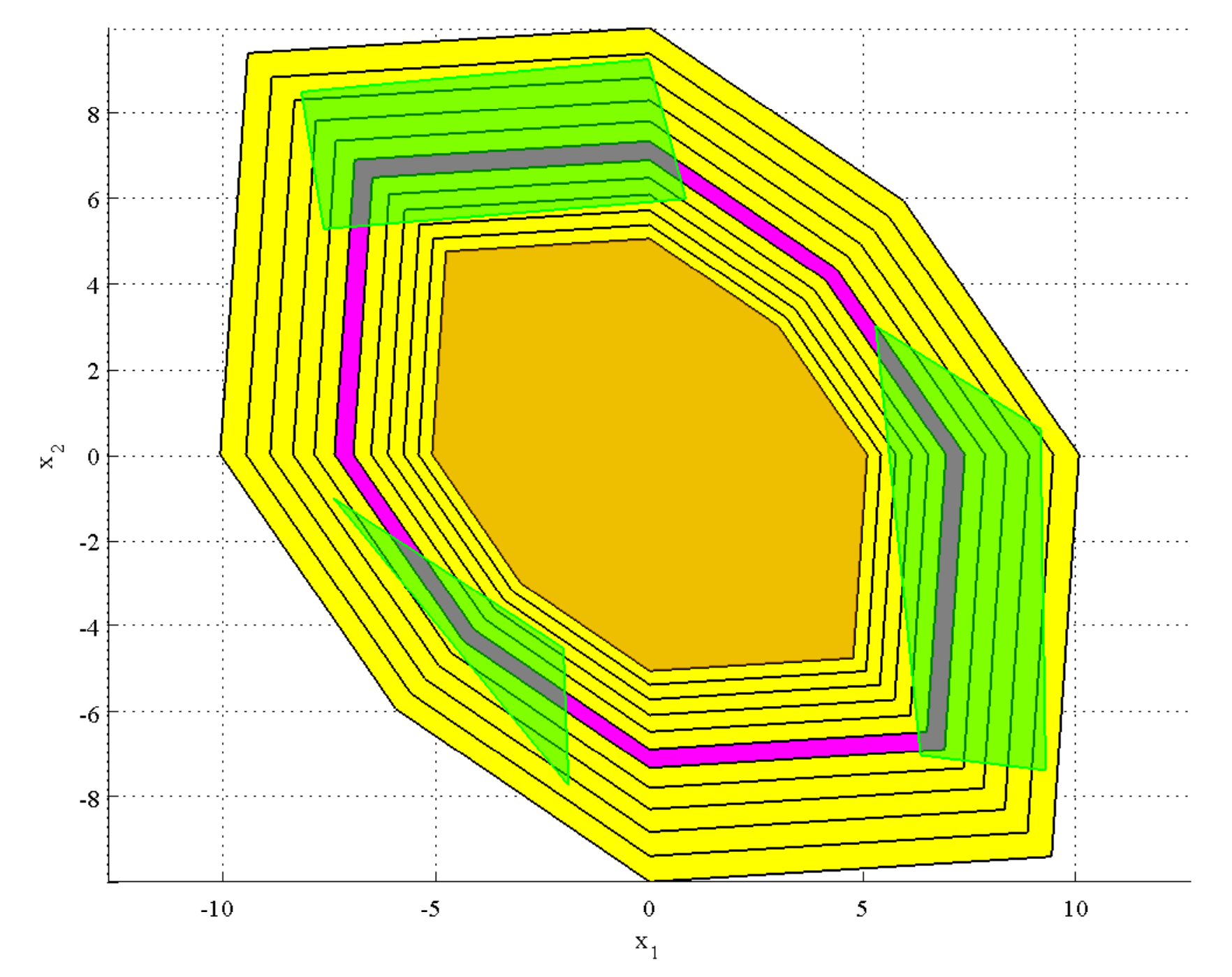}
   \caption{An example of sublevel sets with $N=11$ and one slice $\cS_{6}$ (in purple).}
   \label{fig:levelsets}
\end{figure}
\end{example}

The sublevel sets and the slices are specifically constructed as in \eqref{eq:gammaseq} with the contractive parameter $\rho$, in order to provide the useful property that states within a slice must transition to a lower slice.
\begin{proposition}
\label{prop:slicetrans}
Assume that the set of slices $\{\mc S_{i}\}_{i=0,\ldots,N}$ is obtained by a sequence ${\bar\Gamma}$ satisfying \eqref{eq:gammaseq}.  Given a state $x$ in the $i$-th slice, \ie $x\in \mc S_{i}$, where $1\leq i\leq N$, its successor state ($x'=Ax$) satisfies $x'\in\mc S_{j}$ for some $j<i$.
\end{proposition}
\begin{pf}
From Prop. \ref{prop:rhocontractive}, we have that $\cP_{\Gamma_{i}}$ are $\rho$-contractive.  By the definition of a $\rho$-contractive set (Def. \ref{defPI}), we have that $x'=Ax\in \rho \cP_{i}=\{x\in \rn\st \|Lx\|_{\infty}\leq \rho\Gamma_{i}\}$.   From \eqref{eq:gammaseq}, we have $\rho\Gamma_{i}=\Gamma_{i-1}$.  Therefore $\mc P_{\Gamma_{i-1}}=\{x\in \rn\st \|Lx\|_{\infty}\leq \Gamma_{i-1}\}$ implies that $\mc P_{\Gamma_{i-1}}=\{x\in \rn\st \|Lx\|_{\infty}\leq \rho\Gamma_{i}\}$ and hence  $\mc P_{\Gamma_{i-1}}=\rho \mc P_{\Gamma_{i}}$ and $x'\in\mc P_{\Gamma_{i-1}}$.  From the definition of slices \eqref{eq:slices}, $x'\in S_{j}$ for some $j<i$. \qed
\end{pf}

We now present the abstraction algorithm (see Alg. \ref{alg:main}) that computes the bisimulation quotient.   In Alg. \ref{alg:main}, we make use of two procedures $\mathtt{FindPre}$ and $\mathtt{Refine}$, which will be further explained below.  The main idea is to start with $P_{\mx}$, and iteratively refine the partition until it becomes a refinement to both $P_{\mx}$ as in \eqref{eq:partition} and $\{\mc S_{i}\}_{i=0,\ldots,N}$.  The first procedure is necessary so that the partition is observation preserving.  The second procedure allows us to ensure that at iteration $i$ of the algorithm, the bisimulation quotient for states within $\mc P_{\Gamma_{i}}$ is completed.  Similar to the slices, the solution to Prob.~\ref{prob:main} obtained from Alg.~\ref{alg:main} is a partition consisting of bounded semi-linear sets.

\begin{algorithm}[h]
\caption{Abstraction algorithm}
\begin{algorithmic}[1]
\label{alg:main}
\REQUIRE System dynamics \eqref{eq:linearDyn}, polytopic LF $V(x)=\|Lx\|_{\infty}$ with a contractive rate $\rho$, sets $\mx$, $\mt$ and $\{\mc R_{i}\}_{i\in R}$.
\ENSURE $\mathcal T_{e}\ssim$ as a bisimulation quotient of the embedding transition system $\T_{e}$ and the corresponding observation preserving partition $P$.
\STATE Generate the sequence of sublevel sets $\bar\cP_{\Gamma} = \cP_{\Gamma_{0}},\ldots, \cP_{\Gamma_{N}}$ and slices $\cS_{0},\ldots, \cS_{N}$ as defined in \eqref{eq:slices}.
\STATE Obtain $P_{\mx}$ as in \eqref{eq:partition}.
\STATE Set $P_{0}:=\mathtt{Refine}(P_{\mx}, \{\mc S_{i}\}_{i=0,\ldots,N})$.
\STATE Initialize $\T_{e}/_{\sim_{0}}$ by setting $Q_{e}/_{\sim_{0}}$ as the set labeling $P_{0}$. Set transition only for the state $q\in Q_{e}/_{\sim_{0}}$ where $\eq (q)=\mc S_{0}=\mt$ with $q\to_{\sim_{0}} q$.
\FOR{each $i=0,\ldots,N-1$}
\FOR{each $\tilde P\in P_{i}$ where $\tilde P\subseteq \mc S_{i}$}
\STATE Find $P_{\pre}=\mathtt{FindPre}(\tilde P)$
\STATE Set $P_{i+1}=\mathtt{Refine}(P_{i}, P_{\pre})$.  Update and add the corresponding states in $\T_{e}/_{\sim_{i+1}}$.  Set the transitions of the added states to $\tilde P$ in $\T_{e}/_{\sim_{i+1}}$.
\ENDFOR
\ENDFOR
\STATE Return $\T_{e}/_{\sim_{N}}$ and $P_{N}$ as a solution to Prob. \ref{prob:main}.
\end{algorithmic}
\end{algorithm}
Procedure $\mathtt{FindPre}(\tcp)$ takes as input $\tilde \cP$, a bounded semi-linear set (\eg a slice), and returns the set $\pre_{\T_{e}}(\tilde \cP)$.   In general, the Pre of a semi-linear set is a semi-linear set, and it can be computed via quantifier elimination \citep{bochnak1998real}.  In particular, a bounded semi-linear set $\tcp$ implies that it only belongs to one of the following cases:
\begin{enumerate}
 \item If $\tilde \cP$ is a polytope $\cP$ in the representation $\mc P=\{x\in \rn \st H_{\mc P}x\leq h_{\mc P}\}$ for some $k\geq n+1$, $H_{\mc P}\in \mb R^{k\times n}$ and $h_{\mc P}\in \mb R^{k}$, the $\pre$ of $\mc P$ can be obtained using polytopic operations only, as
\be
\label{eq:prepoly}
\pre_{\T_{e}}(\cP) = \{x\in\rn \st H_{\cP} A x \leq h_{\cP}\},
\ee
which is a possibly degenerate polytope in $\rn$.  Note that \eqref{eq:prepoly} applies to a polytope $\cP$ of any dimension;
\item If $\tcp$ is a union of polytopes, one can use a standard convexation method to decompose $\tcp$ to a set of polytopes $\{\mc P_{i}\}_{i\in I}$ (see, \eg \citep{grunbaum2003convex}).   The $\pre$ of $\tcp$ can then be computed as $\cup_{i\in I}\pre_{\T_{e}}(\mc P_{i})$ using \eqref{eq:prepoly};
\item If $\tcp$ is a convex and bounded semi-linear set, then $\tcp=\cP\setminus \cup_{i\in I}\cP_{i}$ for some polytope $\cP$ and its facet $\cP_{i}\in f(\cP)$.  Since $\T_{e}$ is deterministic, we have $\pre_{\T_{e}}(\tcp)=\pre(\cP)\setminus \pre(\cup_{i\in I}\cP_{i})$, where the second term can be computed as described in case (ii);
\item If $\tcp$ is a general (non-convex) bounded semi-linear set, then again it can be decomposed into convex and bounded semi-linear sets and $\pre_{\T_{e}}(\tcp)$ can be computed as the union of their $\pre$s as described in case (iii).
\end{enumerate}
As summarized above, we see that $\mathtt{FindPre}(\tilde \cP)$ can always be carried out by convex decompositions and repeated applications of \eqref{eq:prepoly}, and thus $\mathtt{FindPre}(\tilde \cP)$ only requires polytopic operations.  Since the $\pre$ of a bounded semi-linear set is a bounded semi-linear set, $\texttt{FindPre}$ can be carried out with polytopic operations throughout Alg. \ref{alg:main}.

The procedure $\mathtt{Refine}(P, \tcp)$ (outlined in Alg. \ref{alg:partition}) refines an observation preserving partition $P$ by partitioning the set $\tcp$, which is assumed to be a bounded semi-linear set\footnote{With a slight abuse of notation, $\mathtt{Refine}(P,\{\tcp\}_{i\in I})$ stands for sequentially applying $\mathtt{Refine}(P, \tcp_{i})$ for each $i\in I$.}.  The proof of correctness of Alg. \ref{alg:partition} is straight-forward, since sets in a partition $P=\{P_{i}\}_{i\in I}$ are piecewise disjoint by definition and as such $\tcp=\bigcup_{i\in I}(P_{i}\cap \tcp)$.  If $P$ consists of bounded semi-linear sets, we can directly see from Alg. \ref{alg:partition} that the resultant refinement $P'$ has the same property.  This fact allows us to use $\mathtt{FindPre}(\tcp)$ for each set $\tcp\in P'$.
\begin{algorithm}[h]
\caption{$P'=\mathtt{Refine}(P, \tcp)$}
\begin{algorithmic}[1]
\label{alg:partition}
\REQUIRE $P$ is an observation preserving partition of $\mx$.  $\tcp\subseteq \mx$ is a bounded semi-linear set.
\ENSURE $P'=\{P'_{i}\}_{i\in I}$ is a finite refinement of $P$, and there exists $J\subseteq I$ such that $\tcp=\cup_{j\in J} P'_{j}$.
\STATE Set $P'=P$
\FORALL{$P'_{i}\in P'$ such that $P'_{i}\cap\tcp\neq \emptyset$}
\STATE Replace $P'_{i}$ in $P'$ by $\{P'_{i}\cap\tcp, P'_{i}\setminus\tcp\}$
\ENDFOR
\end{algorithmic}
\end{algorithm}

The correctness of Alg. \ref{alg:main} will be shown by an inductive argument.  
Given a sublevel set $\mc P_{\Gamma_{i}}$ and a partition $P_{i}$ as obtained in Alg. \ref{alg:main}, we define $\tilde P_{i}$ as 
\be
\label{eq:subpartition}
\tilde P_{i}:=\{P\in P_{i} \st P\subseteq \mc P_{\Gamma_{i}}\}.
\ee  
From Alg. \ref{alg:main}, we see that $P_{0}$ partitions all the slices, and since $P_{i}$ is a finite refinement of $P_{0}$, we can directly see that $\tilde P_{i}$ is a partition of $\mc P_{\Gamma_{i}}$.   We define an embedding transition system $\T_{e}(i)$ as a subset of $\T_{e}$, where its state-space is $\{x\in Q_{e} \st x\in \mc P_{\Gamma_{i}}\}$.  We have the following proposition.
\begin{proposition}
\label{prop:inductionstep}
At the completion of the $i$-th iteration (in the outer loop) of Alg.~\ref{alg:main} (where $P_{i+1}$ is obtained), if $\sim_{i}$ induced by $\tilde P_{i}$ as defined in \eqref{eq:subpartition} is a bisimulation of $\T_{e}(i)$, then $\sim_{i+1}$ induced by $\tilde P_{i+1}$ is a bisimulation of $\T_{e}(i+1)$.
\end{proposition}
\begin{pf}
 If $\sim_{i}$ induced by $P_{i}$ is a bisimulation of $\T_{e}(i)$, then from Prop. \ref{prop:slicetrans}, we have that for each $x\in \mc S_{i+1}$, $x'=Ax$ must be in a lower slice and thus $x'\in\T_{e}(i)$.  For each $x' = Ax$ where $x\in \mc S_{i+1}$, if $x'\in \mc S_{i}$, then we have $x\in P_{\pre}=\mathtt{FindPre}(P)$ (from Step 7 of Alg. \ref{alg:main}) for some $P\in P_{i}$, and after the refinement step (Step 8), we have $x\in P'\subseteq P_{\pre}$ for some $P'\in P_{i+1}$, and $\T_{e}/_{\sim_{i+1}}$ is updated by 1) adding state $\eq(P')$ to $Q_{e}/_{\sim_{i+1}}$ and 2) adding the transition $\eq(P')\to_{\sim_{i+1}} \eq(P)$ .  We note that from the definition of $\pre$, for any $x\in P'$, $x'=Ax\in P$, thus for any $x_{i}\sim x_{j}$, $Ax_{i}\sim Ax_{j}$, and transition $\eq(P')\to_{\sim_{i+1}} \eq(P)$ satisfies the bisimulation requirement.  On the other hand, if $x'\notin S_{i}$, then $x'\in S_{j}$ for some $j<i$ and $x$ is already in a set $P'$ where $\eq(P')\to_{\sim_{i+1}} \eq(P)$ for some $P$ satisfying the bisimulation requirement.  Therefore, step $7$ and $8$ of Alg. \ref{alg:main} provides exactly the transitions needed for states all states in $\mc S_{i+1}$ and thus, $\sim_{i+1}$ induced by $\tilde P_{i+1}$ is a bisimulation of $\T_{e}(i+1)$.
\end{pf}
\begin{proposition}
Alg. \ref{alg:main} returns a solution to Prob.~\ref{prob:main} in finite time.
\end{proposition}
\begin{pf}
From Alg.~\ref{alg:partition}, we have that $P_{i}$ is a refinement of $P_{\mx}$ for any $i=0,\ldots,N$.  Therefore, $P_{N}$ and its induced relation $\sim_{N}$ are observational preserving.

At step $4$ of Alg. \ref{alg:main}, we set $q\to_{\sim_{0}} q$ where $\eq(q)=\mt$.  From the definition of $\T_{e}$, we see that since $\mt$ is the only state, $\sim_{0}$ induced by $\tilde P_{0}$ is a bisimulation of $\T_{e}(0)$.  Using Prop. \ref{prop:inductionstep} and induction, at iteration $N-1$, we have that $\sim_{N}$ induced by $\tilde P_{N}$ is a bisimulation of $\T_{e}(N)$.   Note that $\tilde P_{N}$ is exactly $P_{N}$, $\mc P_{\Gamma_{N}}$ is exactly $\mx$ and $\T_{e}(N)$ is exactly $\T_{e}$.  Therefore $\sim_{N}$ induced by $P_{N}$ is a bisimulation of $\T_{e}$.

Finally, note that at each iteration, the number of sets updated are finite.  Therefore, the bisimulation quotient is finite and moreover Alg.~\ref{alg:main} completes in finite time.   \qed
\end{pf}

\begin{figure*}[ht]
	\centering
	\subfloat[]{\includegraphics[scale=.28]{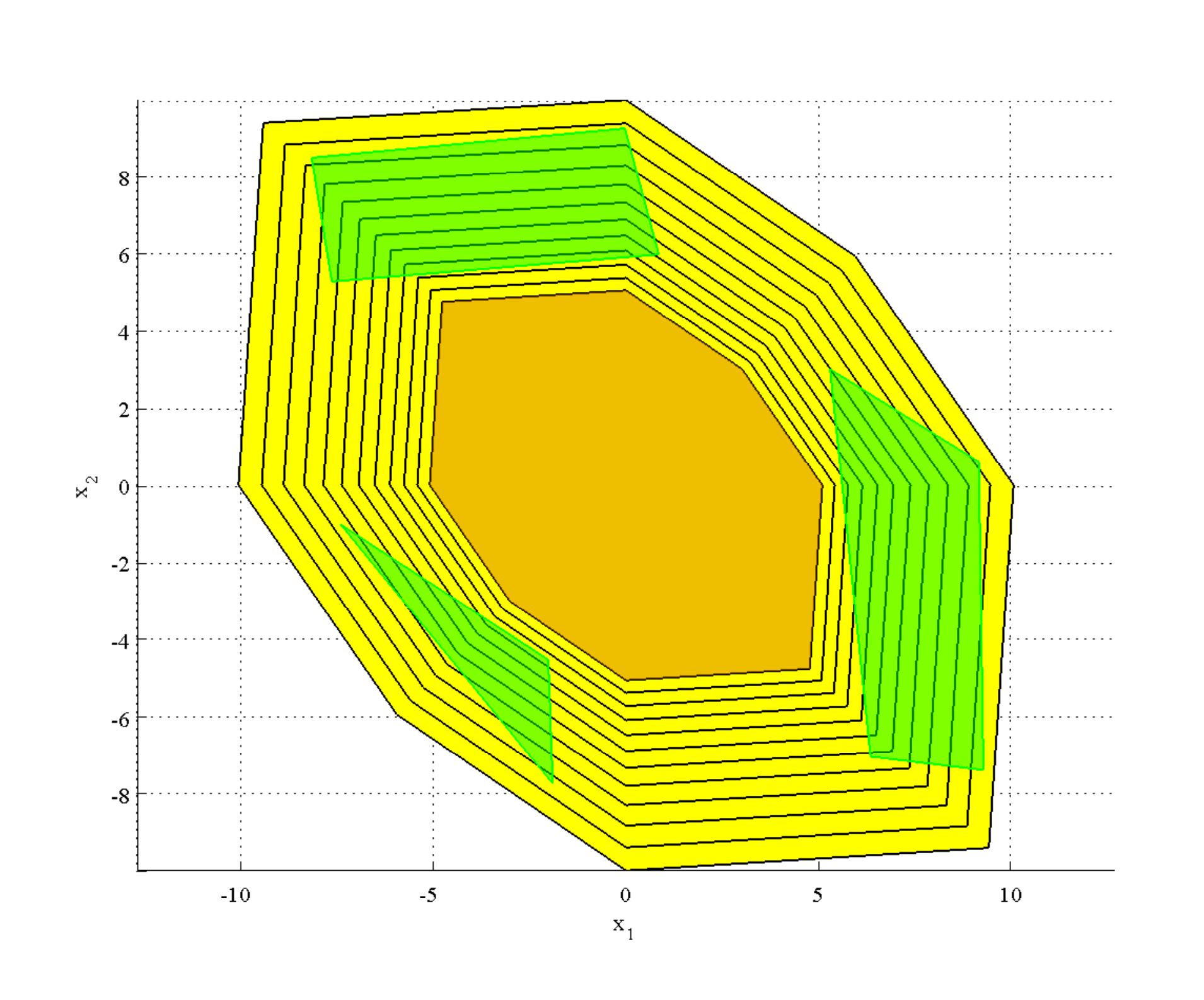} \label{fig:seq:initnoslice}}
	\subfloat[]{\includegraphics[scale=.28]{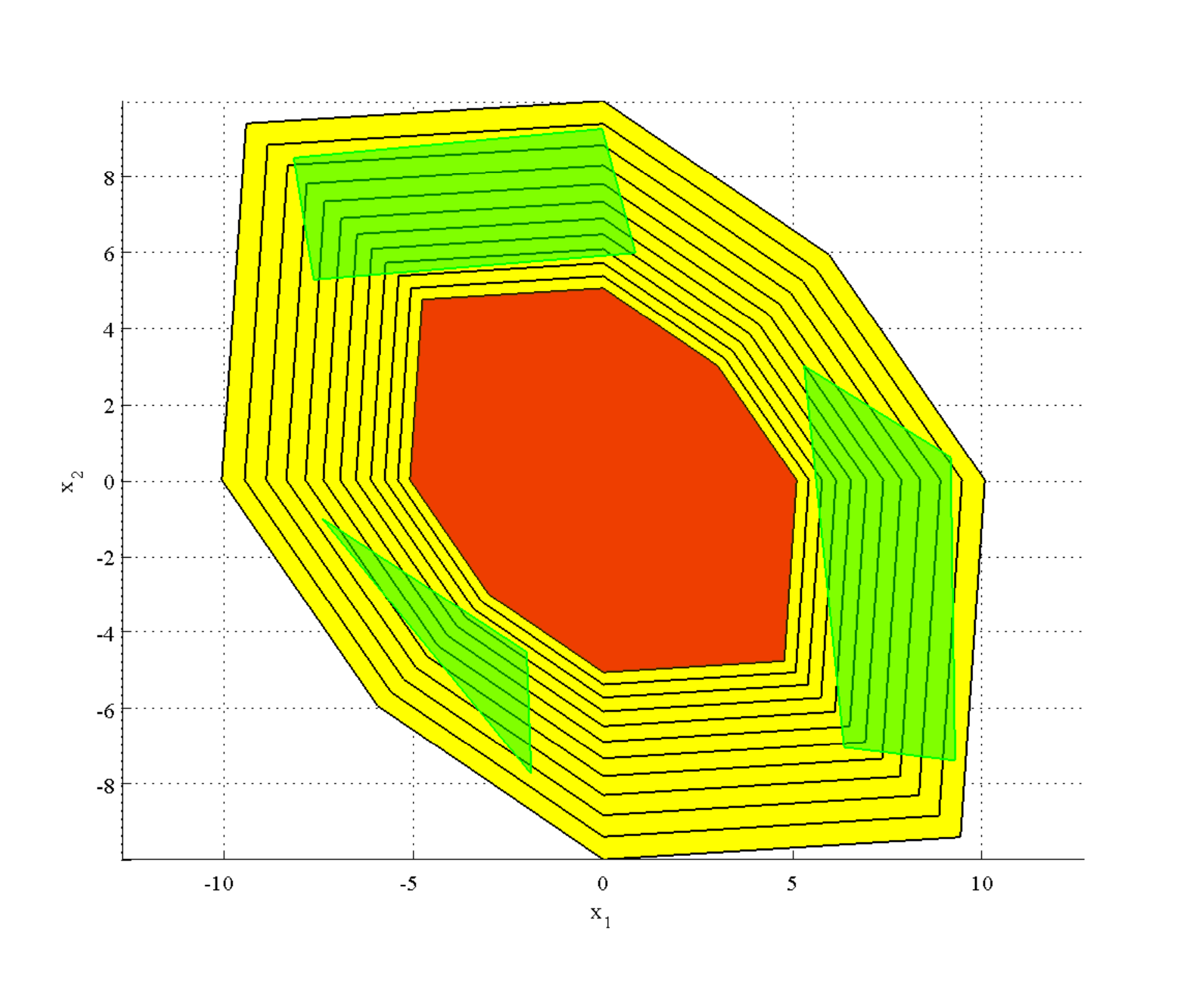}}	
	\subfloat[]{\includegraphics[scale=.28]{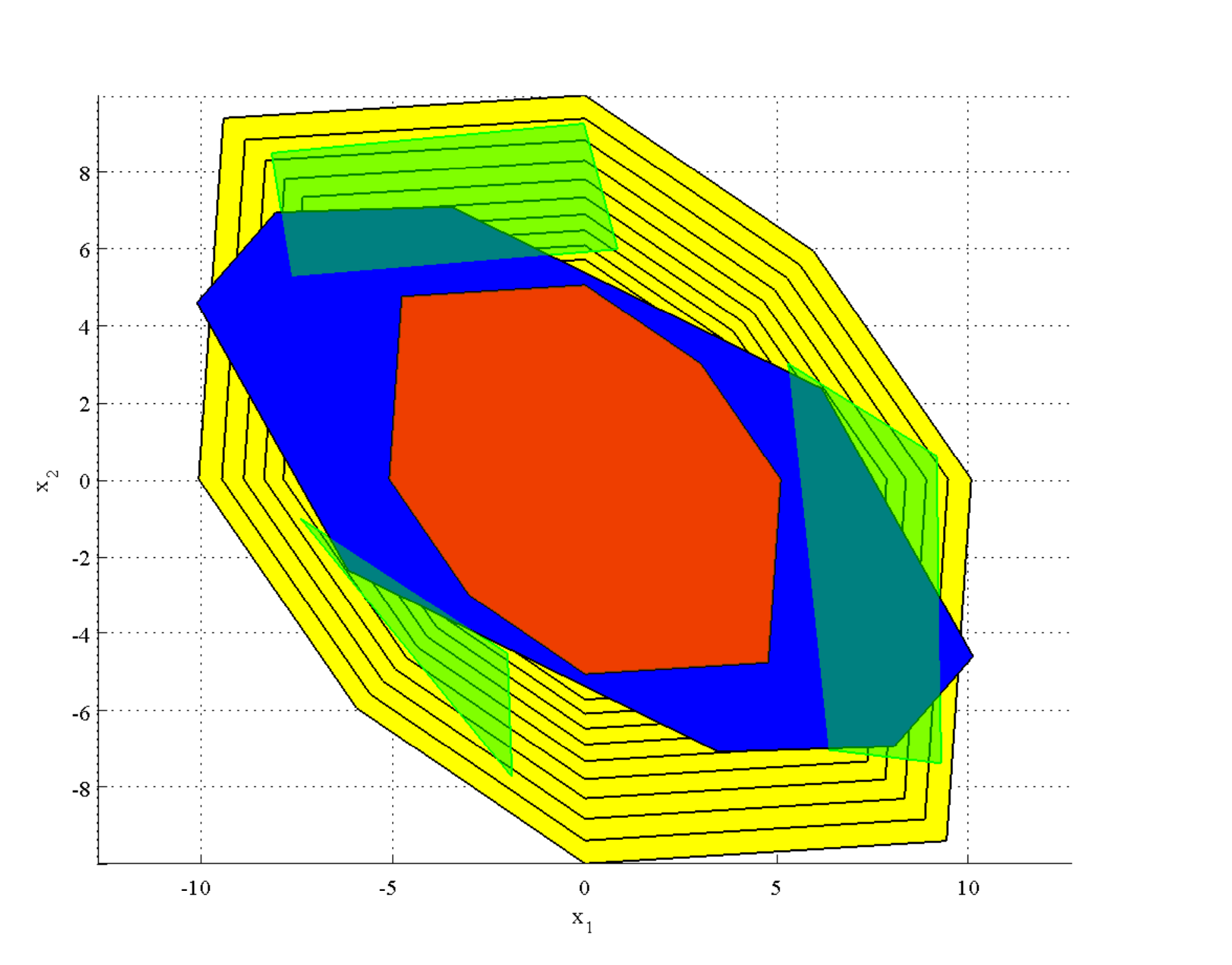}}\\
	\subfloat[]{\includegraphics[scale=.28]{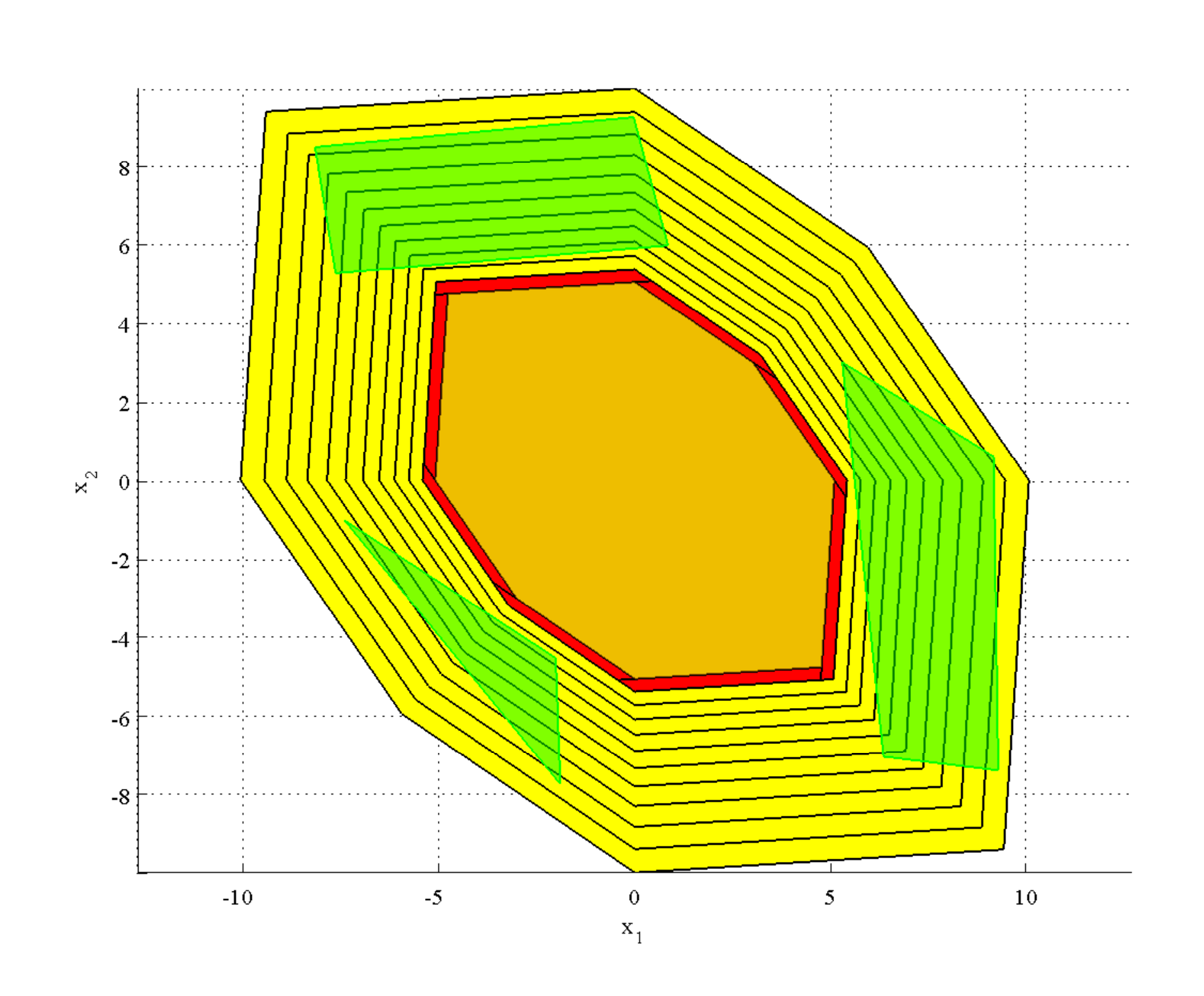}}
	\subfloat[]{\includegraphics[scale=.28]{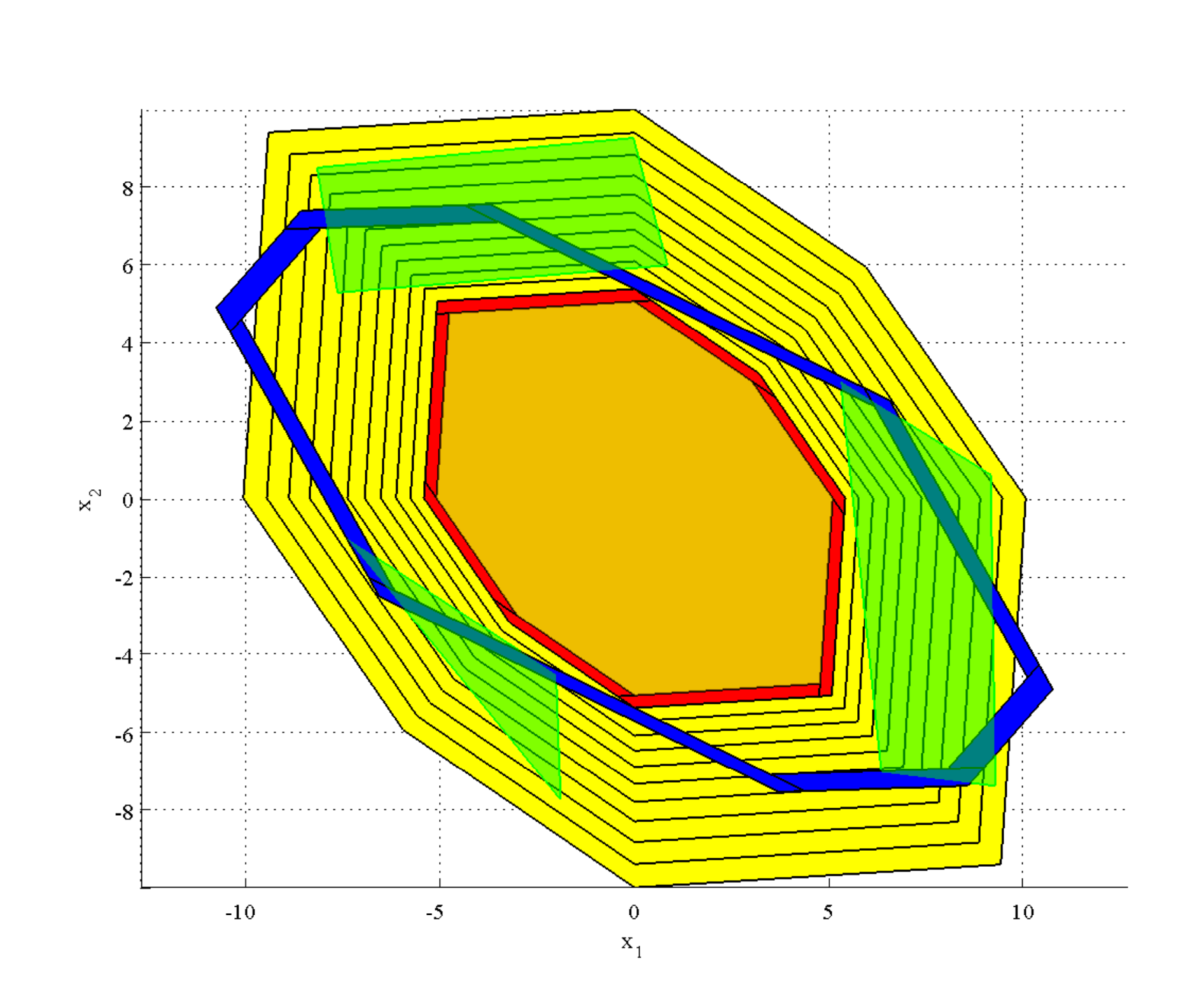}}
	\subfloat[]{\includegraphics[scale=.28]{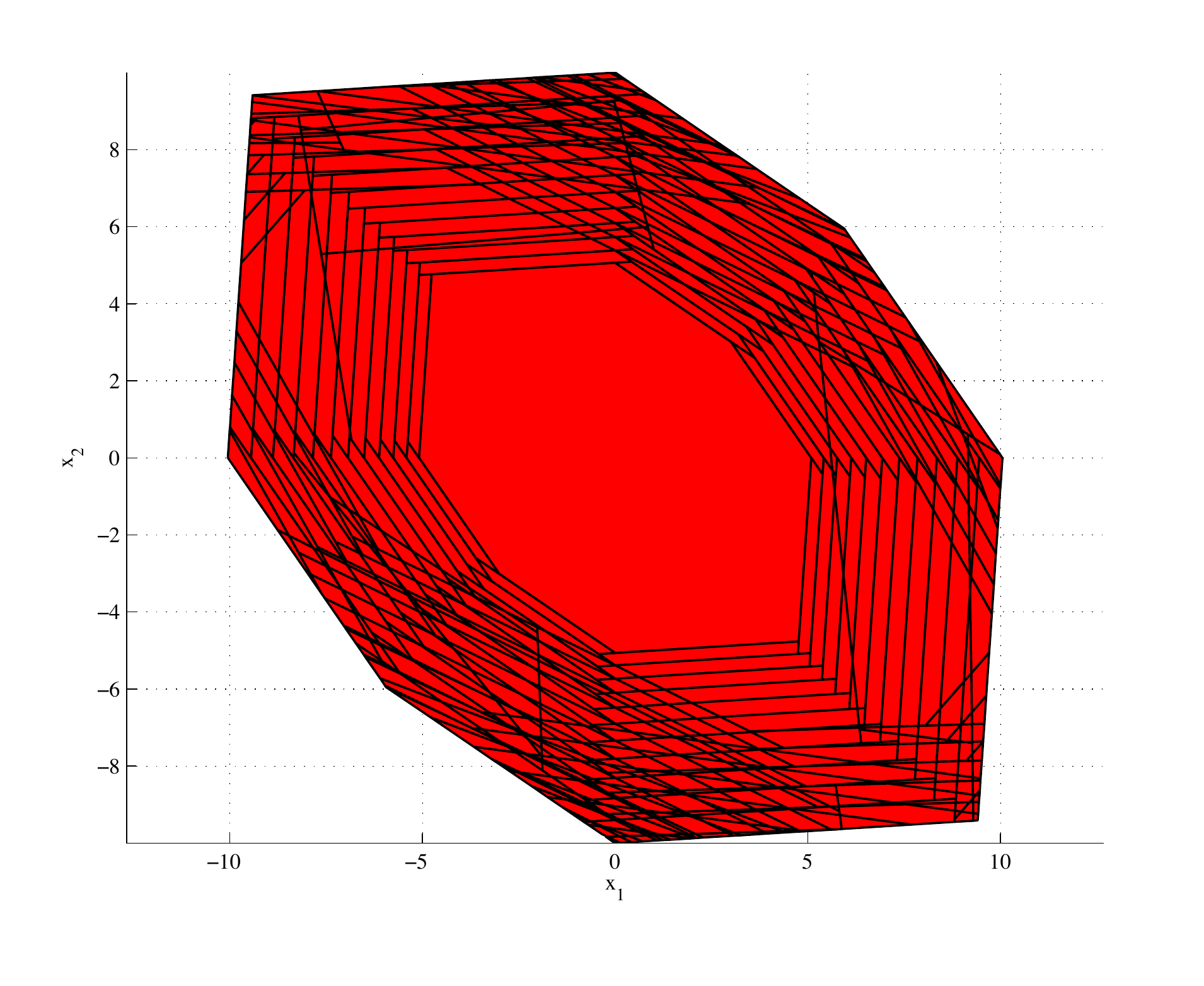}}
\caption{(a) The slices are shown in yellow, except $\mt$, which is shown in light brown.  The observed regions are shown in transparent green.  (b) In the first iteration ($i=0$), the slice $\cS_{0}=\mt$ is shown in red.  (c) The $\pre$ of $\cS_{0}$ is shown in blue.  At this point, the bisimulation quotient for states within $P_{\Gamma_{0}}=\mt$ is completed, which consists of just a single state.  (d) In the second iteration ($i=1$), the slice $\cS_{1}$ is shown in red.  (e) The $\pre$ of $\cS_{1}$ is shown in blue.  At this point, the bisimulation quotient for states within $P_{\Gamma_{1}}$ is completed.  (f)  At the last iteration where $i=10$, the algorithm is completed.  The state space covered by the bisimulation quotient is shown in red, covering all of $\mx$.}
\label{fig:snapshots}
\end{figure*}

\begin{example}[Example \ref{ex:step2} continued]
 \label{ex:step3}
Alg.~\ref{alg:main} is applied on the same setting as in Example~\ref{ex:step2} to computate the bisimulation quotient.  ``Snapshots'' of the algorithm iterations are shown in Fig.~\ref{fig:snapshots}.  The final result is a Transition system with $320$ states.  In this example, Alg.~\ref{alg:main} was completed in 3 minutes on a Macbook Pro 2011 model.
\end{example}

\section{System verification with Linear Temporal Logic formulas}
\label{sec:verification}
In this section we show how we can use the bisimulation quotient obtained as a solution to Prob.~\ref{prob:main} to verify the behavior of system \eqref{eq:linearDyn} in the state space $\mx\setminus\mt$ over the observed regions $\{\mr_i\}_{i\in R}$ and the observation $\Pi_\mt$ corresponding to $\mt$.  We will employ Linear Temporal Logic (LTL) to describe high level system specifications. A detailed description of the syntax and semantics of LTL is beyond the scope of this paper and can be found in, for example, \citep{Clarke99}.  Roughly, an LTL formula is built up from a set of atomic propositions $\Pi$, which are properties that can be either true or false, standard Boolean operators $\neg$ (negation), $\Lor$ (disjunction), $\Land$ (conjunction), and temporal operators $\Next$ (next), $\Until$ (until), $\Event$ (eventually), $\Always$ (always) and $\Rightarrow$ (implication). The semantics of LTL formulas are given over words, which is defined as an infinite sequence $\bo=o_{0}o_{1}\ldots$, where $o_{i} \in 2^{\Pi}$ for all $i$.  We say $\bo\vDash\phi$ if the word $\bo$ satisfies the LTL formula $\phi$.  We say a trajectory $\bq$ of a transition system $\T$ satisfies LTL formula $\phi$, if the word generated by $\T$ (see Def. \ref{def:tran_sys}) satisfies $\phi$.


\begin{example}\label{ex:LTLexample}
Again, consider the setting in Example \ref{ex:simpleEx} with $\mc R=\{\mr_i\}_{i=\{1,2,3\}}$.  We now consider a specification in LTL over $R=\{1,2\}$.  For example, the specification:

\emph{``The system trajectory never visits Region $2$ and eventually visits Region $1$.  Moreover, if it visits Region $3$ then it must not visit Region $1$ at the next consecutive time instant''}

can be translated to an LTL formula:
\be
\label{eq:exampleLTLformula}
\phi:=\Always \neg 2 \Land \Event 1 \Land (3 \Rightarrow \Next \neg 1)
\ee
\end{example}

\begin{remark}
Set $\mt$ is by definition positively invariant. Therefore, all trajectories of \eqref{eq:linearDyn} eventually reach $\mt$.  As a result, we see that any LTL formula satisfiable by \eqref{eq:linearDyn} must not violate formula $\Event \Pi_\mt$.  For example $\psi=\Always \neg \Pi_\mt \Land \phi$ is not satisfiable by the system for any LTL formula $\phi$ as the first part of $\psi$ is in contradiction to $\Event \Pi_\mt$.
\end{remark}

\begin{problem}
\label{prob:LTL}
Let system \eqref{eq:linearDyn} with a polyhedral Lyapunov function in the form of \eqref{eq:polyLF}, sets $\mx$, $\mt$ and $\{\mr_i\}_{i\in R}$, and an LTL formula $\phi$ over $R\cup \Pi_\mt$ be given.  
Find the largest set $\mc S\subseteq Q_{e}$ such that state trajectories of the embedding transition system $\T_{\e}$  originating from $\mc S$ satisfy $\phi$.
\end{problem}
Our solution to Prob. \ref{prob:LTL} proceed by finding a bisimulation quotient $\T/_\sim$ of the embedding transition system $\T_e$ using Alg. \ref{alg:main}.  Then we translate $\phi$ to a so-called B\"{u}chi Automaton, defined below.
\begin{definition}\label{def:omega_aut}
A (non-deterministic) B\"{u}chi automaton is a tuple $\B=(S_{\B},S_{\B0},\Sigma,\delta,F_{\B})$, where
\begin{itemize}
\item $S_{\B}$ is a finite set of states;
\item $S_{\B0}\subseteq S_{\B}$ is the set of initial states;
\item $\Sigma$ is the input alphabet;
\item $\delta: S_{\B}\times \Sigma \rightarrow 2^{S_{\B}} $ is the transition function;
\item $F_{\B}\subseteq S$ is the set of accepting states.
\end{itemize}
We denote $s\overset{\sigma}{\to}_{\B}s'$ if $s'\in \delta(s,\sigma)$.   A word $\sigma_{0}\sigma_{1}\ldots$ over $\Sigma$ generates trajectories $s_{0}s_{1}\ldots$ where $s_{0}\in S_{\B0}$ and $s_{k}\overset{\sigma_{k}}{\to}_{\B}s_{k+1}$ for all $k\geq 0$.  $\B$ accepts a word over $\Sigma$ if it generates at least one trajectory on $\B$ that intersects $F_{\mathcal B}$ infinitely many times.
\end{definition}
For any LTL formula $\phi$ over $\Pi$, one can construct a B\"{u}chi automaton with input alphabet $\Sigma= 2^{\Pi}$ accepting all and only words over $2^{\Pi}$ satisfying $\phi$ \citep{Clarke99}.    Algorithms and implementations for the translation from $\phi$ to a corresponding B\"{u}chi automaton $\mathcal B$ can be found in \citep{gastin2001fast}.
\begin{definition}
\label{def:PA}
 Given a transition system $\T=(Q,\to,\Pi,h)$ and a B\"uchi automaton $\B = (S_{\B},S_{\B 0},2^{\Pi},\delta_\B,F_{\B})$, their product automaton, denoted by $\Prod=\T \times \B$, is a tuple $\Prod=(S_\Prod,S_{\Prod0},\Delta_\Prod,F_\Prod)$
where
\begin{itemize}
 \item $S_\Prod = Q \times S_{\B}$;
 \item $S_{\Prod0} = Q \times S_{\B 0}$;
 \item $\Delta_{\Prod}\subseteq S_{\Prod}\times S_{\Prod}$ is the set of transitions, defined by: $\left((q,s),(q',s')\right)\in \Delta_{\Prod}$ iff $q\to q'$ and $s\overset{h(q)}{\longrightarrow}_{\B}s'$;
 \item $F_\Prod = Q \times F_{\B}$.
\end{itemize}
We denote $(q,s) \to_{\Prod} (q',s')$ if $((q,s), (q',s'))\in \Delta_{\Prod}$.
A trajectory ${\rm \bp}=(q_{0}, s_{0})(q_{1},s_{1})\ldots$ of $\Prod$ is an infinite sequence such that $(q_{0}, s_{0})\in S_{\Prod 0}$ and $(q_{k},s_{k}) \to_{\Prod} (q_{k+1},s_{k+1})$ for all $k\geq 0$.   Trajectory $\bp$ is called accepting if and only if it intersects  $F_{\Prod}$ infinitely many times.
\end{definition}

By the construction of $\Prod$ from $\T$ and $\B$, $\bp$ is accepted if and only if $\bq=\gamma_{\mathcal T}(\bp)$ satisfies the LTL formula corresponding to $\B$ \citep{Clarke99}, where $\gamma_{\mathcal T}(\bp)$ is the projection of a trajectory $\bp$ on $\mc P$ onto $\mc T$ by simply removing the automaton part of the state in $(q,s)\in S_{\Prod}$.

\begin{remark}
Normally the product automaton is constructed from a transition system with an initial state $q_{0}$, whereas the transition system generated as a solution to Prob. \ref{prob:main} is not initialized.  Since any state $q\in Q_{e}\ssim$ can be an initial condition, the set of initial states of $\Prod$ is $Q_{e}\ssim \times S_{\B0}$.   Thus, here we augment the definition of $\Prod$ slightly so that it is constructed as a product of an uninitialized transition system and a B\"uchi automaton.
\end{remark}

In \citep{ding2010receding}, an algorithm was proposed to compute the largest subset $F^\star_\Prod\subseteq F_\Prod$ such that it can reach another state in $F^\star_\Prod$.  The following property was shown to hold:

\begin{proposition}
\label{prop:cdcresult}
A trajectory $\bp$ is accepting if and only if each accepting state appearing in $\bp$ is in $F^\star_\Prod$.
\end{proposition}
A state $q\in Q$ of $\T$ from which the trajectory satisfies the formula must be such that a state in $F^\star_\Prod$ is reachable from $(q,s_0)$ for some $s_0\in S_{\mc B0}$.  Therefore, Prob. \ref{prob:LTL} can be solved by a simple reachability analysis for the set $F^\star_\Prod$ on the product automaton.  Note that during the generation of set $F^\star_\Prod$ in the algorithm proposed in \citep{ding2010receding}, the reachability is already determined for each state in $\Prod$, so no extra computation is necessary.  This procedure is summarized in the following algorithm.

\begin{algorithm}[h]
\caption{Finding the largest subset satisfying an LTL formula}
\begin{algorithmic}[1]
\label{alg:LTL}
\REQUIRE $\mx$, $\mt$, $\{\mr_i\}_{i\in R}$, and an LTL formula $\phi$ over $R\cup \Pi_\mt$
\ENSURE The largest set $\mc S\subseteq Q_{e}$, such that the embedding transition system $\T_{e}$ with the initial state $q_0\in \mc S$ produces a word satisfying $\phi$
\STATE Generate the bisimulation quotient $\T_{e}/_\sim$ for $\T_e$.
\STATE Translate $\phi$ to a Buchi automaton $\mc B$
\STATE Generate the product $\Prod$ between $\T_{e}/_\sim$ and $\mc B$
\STATE Find the subset $F^\star_\Prod\subseteq F_\Prod$ with the algorithm by \cite{ding2010receding}.
\STATE $\mc S=\{\eq(q) \st q \in \T_{e}/_\sim \textrm{ and there exists } s_0\in S_{\mc B0} \textrm{ such that } F^\star_\Prod(q,s_0) $ $\textrm{is reachable from } (q,s_0)\}$
\end{algorithmic}
\end{algorithm}
\begin{proposition}
 Upon termination, Alg.~\ref{alg:LTL} gives a solution to Prob.~\ref{prob:LTL}.
\end{proposition}
\begin{pf}
We prove that Alg.~\ref{alg:LTL} generates the largest set of satisfying states by contradiction. From the last step of Alg.~\ref{alg:LTL}, we have that $\mc S=\{\eq(q) \st q \in \T_{e}/_\sim \textrm{ and } \exists s_0\in S_{\mc B0} \textrm{ such that } F^\star_\Prod(q,s_0) \textrm{ is reachable from } (q,s_0)\}$.  Assume that there exists $q_{e}\notin \mc S$ such that a trajectory from $q_{e}$ satisfies $\phi$, and $q_{e}\in eq(q)$ where $q\in Q_{e}\ssim$. In this case, on the product $\T_{e}\ssim\times \B$, from a state $(q,s_{0})\in S_{\Prod 0}$, a state in $F^{\star}_{\Prod}$ cannot be reached, and from Prop. \ref{prop:cdcresult}, we have that trajectory $\bp$ cannot be accepting on $\T_{e}\ssim\times \B$ and $\gamma_{\T_{e}\ssim}(\bp)$ as a trajectory of $\T_{e}\ssim$ cannot be accepting.  Therefore, $\mc L_{\T_{e}\ssim}(q)$ does not satisfy $\phi$.   By the property of language equivalence of bisimulations, we have $\mc L_{\T_{e}}(q_{e})\subseteq \mc L_{\T_{e}}(eq(q))=\mc L_{\T_{e}\ssim}(q)$, and therefore the trajectory from $q_{e}$ cannot be accepting, which violates the above assumption. \qed
\end{pf}

\begin{example}[Example \ref{ex:LTLexample} continued]
For the example specification $\phi$ as in \eqref{eq:exampleLTLformula}, we obtained the solution to Prob. \ref{prob:LTL} by following Alg. \ref{alg:LTL}.  The set of initial states from which the state trajectories satisfy \eqref{eq:exampleLTLformula} are shown in Fig. \ref{fig:ltlresult}.
\begin{figure}[h]
   \center
   \includegraphics[scale=.38]{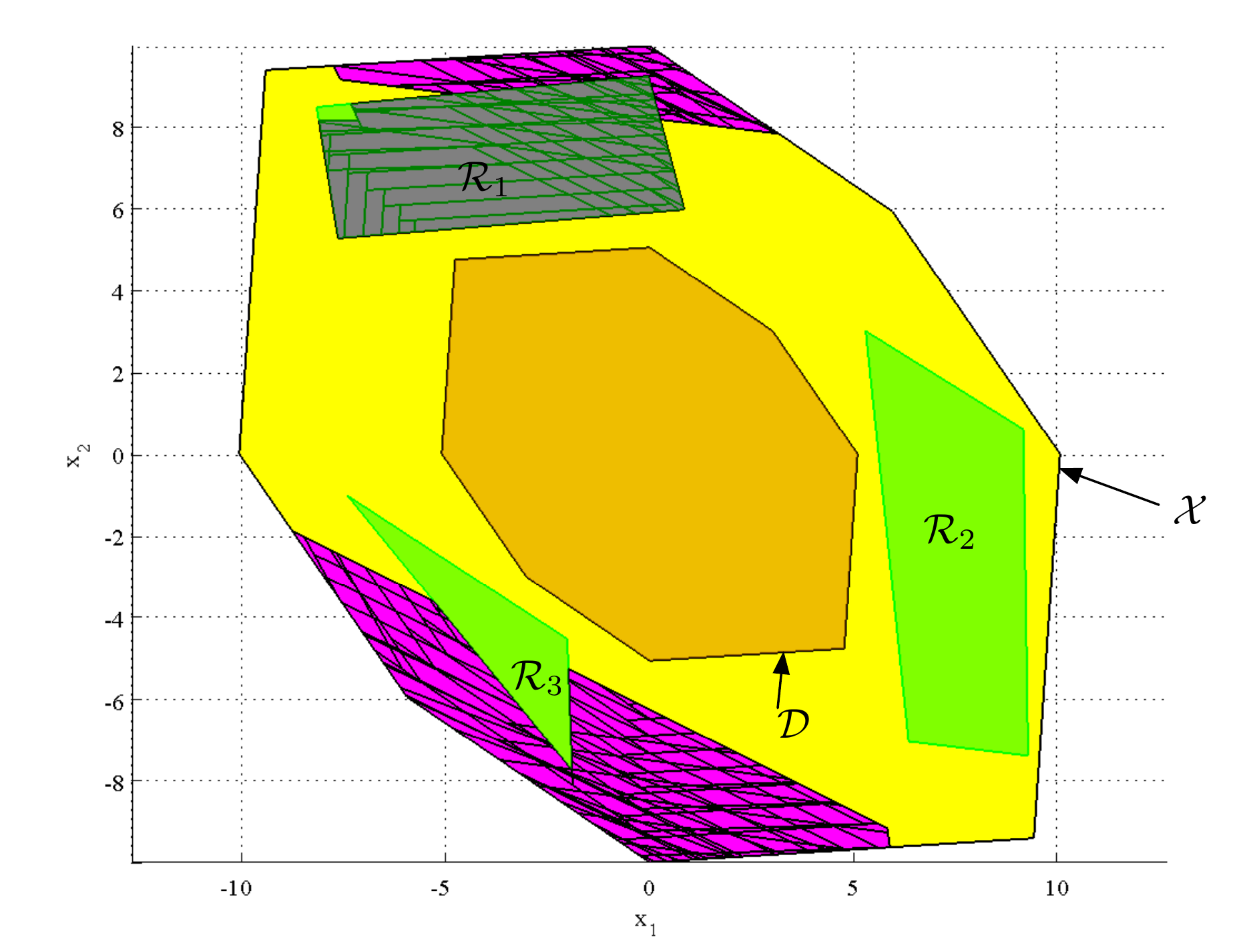}
   \caption{The set of states satisfying $\phi$ (in purple).}
   \label{fig:ltlresult}
\end{figure}
\end{example}


\section{Conclusions and final remarks}
\label{sec:concl}
In this paper we presented a method to abstract the behavior of an autonomous linear system within a positively invariant subset of $\Rset^n$ to a finite transition system via bisimulation.  We employed polyhedral Lyapunov functions to guide the partitioning of the state space and showed that this results requires only polytopic operations.

Future work deals with an extension to continuous-time linear systems and other classes of systems that admit polyhedral Lyapunov functions, in particular, switched linear systems.  We also aim to relax some assumptions and improve the computational complexity of the approach by reducing the size of the bisimulation quotient.
\bibliography{Papers}
\end{document}